\documentclass[10pt]{article}
\usepackage{amsfonts}
\usepackage{amsmath}
\usepackage{rotating}

\csname @addtoreset\endcsname{equation}{section}
\newcommand{\eq}{\begin{equation}}
\newcommand{\feq}{\end{equation}}
\newcommand{\eqn}{\begin{eqnarray}}
\newcommand{\feqn}{\end{eqnarray}}
\newcommand{\arr}{\begin{eqnarray*}}
\newcommand{\farr}{\end{eqnarray*}}

\begin{document}

\begin{titlepage}
\begin{flushright}
IFUM-830-FT
\end{flushright}
\vskip 10mm
\begin{center}
\renewcommand{\thefootnote}{\fnsymbol{footnote}}
{\Large \bf A simple parametrization for \boldmath{$G_2$}}
\vskip 12mm
{\large \bf {Sergio L.~Cacciatori$^{1,2}$\footnote{sergio.cacciatori@mi.infn.it}
}}\\
\renewcommand{\thefootnote}{\arabic{footnote}}
\setcounter{footnote}{0}
\vskip 5mm
{\small
$^1$ Dipartimento di Matematica dell'Universit\`a di Milano,\\
Via Saldini 50, I-20133 Milano, Italy. \\

\vspace*{0.4cm}

$^2$ INFN, Sezione di Milano,\\
Via Celoria 16,
I-20133 Milano, Italy.\\

}
\end{center}
\vspace{2cm}
\begin{center}
{\bf Abstract}
\end{center}
{\small 
We give a simple parametrization of the $G_2$ group, which is consistent with
the structure of $G_2$ as a $SU(3)$ fibration. We also explicitly compute the (bi)invariant
measure, which turns out to have a simple expression.
}

\end{titlepage}

%%%%%%%%%%%%%%%%%%%%%%%%%%%%%%%%%%%%%%%%%%%%%%%%%%%%%%%%%%%%%%%%%%%%%%%%%%%%%%%%%%%%%%%%%%%%%%%%%
\section{Introduction}
\label{intro}
Group theory plays an important role in physics. When a group appears as a symmetry, as for example a
gauge symmetry, one needs to integrate over the group; to have an unbroken symmetry in a path-integral
formulation of the theory, the integration measure must be invariant under the group action.\\
For semisimple groups there is a unique (up to normalization constants) invariant measure, obtained pulling back
the Killing form on the algebra  via the left action of the group over itself. In general however difficulties
arise in finding an explicit parametrization of the group elements which give a simple expression for the
invariant measure and also permit a full determination of the range of parameters. This is in fact what one
needs to do explicit computations.\\
\indent In \cite{Cacciatori:2005yb}, a solution of this problem was found for the group $G_2$. There the group
parametrization emphasizes the structure of the group as a fibration with $SO(4)$ as fiber and the space
$\mathcal{H}$ of quaternionic linear subalgebras of octonions as base. Here I give another parametrization
for $G_2$, based on the well known fact that $G_2$ can be seen as an $SU(3)$ fibration over the six sphere 
$S^6$.\footnote{See for example \cite{Adams}.} \\
The resulting measure turn out to have a very simple expression, with all the parameters varying
in a $14-$dimensional hypercube.  

%%%%%%%%%%%%%%%%%%%%%%%%%%%%%%%%%%%%%%%%%%%%%%%%%%%%%%%%%%%%%%%%%%%%%%%%%%%%%%%%%%%%%%%%%%%%%%%%%%
%\section{Description of the group}
%\label{sec:group}

%%%%%%%%%%%%%%%%%%%%%%%%%%%%%%%%%%%%%%%%%%%%%%%%%%%%%%%%%%%%%%%%%%%%%%%%%%%%%%%%%%%%%%%%%%%%%%%%%%^%%%%%%%%%%%%%%
\section{The representation algebra and the group ansatz}
\label{sec:ansatz}
For the algebra we choose the fundamental ({\bf 7}) representation as in
\cite{Cacciatori:2005yb}. We repeat in appendix \ref{app:comm} only the commutators matrix $B_{IJ} := [C_I , C_J]$.
Now note that $\{ C_i \}_{i=1}^8$ generate the $su(3)$ algebra. Thus we want
to use this fact to construct a new parametrization for the $G_2$ elements,
which underlines the $SU(3)$ subgroup.\\
To this end let us note that $C_9$ commutators with $su(3)$ generate all the
remaining generators of the $g_2$ algebra. We can then hope to find a parametrization
of the group miming the Euler parametrization for $SU(n)$. That is we write
the generic element of $G_2$ in the form
\eqn
g=SU(3)[\alpha_1, \ldots ,\alpha_8] e^{\beta C_9} SU(3)
[\gamma_1 \ , \ldots \ ,\gamma_8]      \label{redundant}
\feqn
where $SU(3)[\alpha_1, \ldots ,\alpha_8]$ is the generic Euler parametrization
of $SU(3)$ shown in appendix \ref{app:su(3)}. \\
However here we have three redundant parameters, the dimension of $G_2$ being $14$.
We can easily eliminate it as follows. In (\ref{redundant}) the left $SU(3)$
term has the form
\eqn
SU(3)[\alpha_1, \ldots ,\alpha_8] =h[\alpha_1, \ldots ,\alpha_5]
e^{\alpha_6 C_3} e^{\alpha_7 C_2} e^{\alpha_8 C_3} \ .
\feqn
Now ${C_1 \ , C_2 \ , C_3}$ commute with $C_9$ so that we can absorb
${\alpha_1 \ , \alpha_2 \ , \alpha_3}$ in the right $SU(3)$ in (\ref{redundant}).\\
We finally obtain the ansatz
\eqn
g[\alpha_1 \ , \ldots \ , \alpha_6 \ ; \gamma_1 \ , \ldots \ , \gamma_8] =
\Sigma[ \alpha_1 \ , \ldots \ , \alpha_6 ]
SU(3)[\gamma_1 \ , \ldots \ ,\gamma_8]
\label{ansatz}
\feqn
with
\eqn
\Sigma[ \alpha_1 \ , \ldots \ , \alpha_6 ]=
e^{\alpha_1 C_3} e^{\alpha_2 C_2} e^{\alpha_3 C_3} e^{\frac {\sqrt 3}2 \alpha_4 C_8}
e^{\alpha_5 C_5} e^{\frac {\sqrt 3}2 \alpha_6 C_9}
\feqn
To show that this ansatz solves our problem we must show that we can choose
the range of the parameters such to cover one time all $G_2$ (up to a subset
of vanishing measure).
%%%%%%%%%%%%%%%%%%%%%%%%%%%%%%%%%%%%%%%%%%%%%%%%%%%%%%%%%%%%%%%%%%%%%%%%%%%%%%%%%%%%%%%%%%
\subsection{Fibration and computation of the metric}
\label{sec:fibre}
We said that the quotient of $G_2$ with the
$SU(3)$ subgroup is a sphere $S^6$. We will use this information to establish
the range of the parameters. For brevity we will call
$S:=SU(3)[\gamma_1 \ , \ldots \ , \gamma_8]$.\\
The metric on the group can be obtained from the left invariant currents
\eqn
Jg =g^{-1} dg = \sum_{I=1}^{14} Jg^I C_I
\feqn
via\footnote{ The factor $-\frac 14$ is due to the normalization
$Trace \{ C_I C_J \} = -4 \delta_{IJ}$. Choosing a metric $(C_I | C_J) =\delta_{IJ}$ on the
algebra is equivalent to fix normalizations such that long and short roots of $g_2$ have length
$2$ and $\frac 2{\sqrt 3}$ respectively \cite{Cacciatori:2005yb}.}
\eqn
ds^2_{G_2} =-\frac 14 Trace\{ Jg \otimes Jg\} \ .
\feqn
If we now write
\eqn
&& J_{\Sigma} = \Sigma^{-1} \cdot d\Sigma =\sum_{I=1}^{14} J_{\Sigma}^I C_I \ , \\
&& J_S =dS \cdot S =\sum_{I=1}^{8} J_S^I C_I \ ,
\feqn
it is straightforward to show that
\eqn
ds^2_{G_2} =\sum_{I=1}^8 \left( J_S^I +J_{\Sigma}^I \right)^2
+\sum_{I=9}^{14} \left( J_{\Sigma}^I \right)^2 \ . \label{metric}
\feqn
Now $\Sigma$ parametrizes the quotient space between $G_2$ and the $SU(3)$
orbits generated by $S$. Fixing the base point $\Sigma$ we recover the $SU(3)$
invariant metric
\eqn
ds^2_{SU(3)} =\sum_{I=1}^8 \left( J_S^I \right)^2 \ .
\feqn
To find the induced metric on the base one must compute the current
$\Sigma^{-1} \cdot d\Sigma $ and project out the components tangent
to the $su(3)$ directions. Doing so one finds
\eqn
ds^2_{BASE} = \sum_{I=9}^{14} \left( J_{\Sigma}^I \right)^2 \ .
\label{S6}
\feqn
If all goes right this must be the metric of a six sphere. We now show that this
is exactly the case. \\
Using the results shown in appendix \ref{app:curr} for the currents, one finds
\eqn
\frac 43 ds^2_{BASE} =d\alpha_6^2 +\sin^2 \alpha_6
\left\{ d\alpha_5^2 +\cos^2 \alpha_5 d\alpha_4^2 +\sin^2 \alpha_5 \left[
s_1^2 +s_2^2 +\left( s_3+\frac 12 d\alpha_4 \right)^2 \right] \right\} \label{6metric}
\feqn
where
\eqn
&& s_1 =-\sin (2\alpha_2) \cos (2\alpha_3) d\alpha_1 +\sin(2\alpha_3)d\alpha_2 \cr
&& s_2 =\sin (2\alpha_2) \sin (2\alpha_3) d\alpha_1 +\cos(2\alpha_3)d\alpha_2 \cr
&& s_3 =\cos (2\alpha_2) d\alpha_1 +d\alpha_3
\feqn
We can now recognize the metric of a six sphere $S^6$ with coordinates
$(\alpha_6\ , \vec X )$, where $\alpha_6$ is the azimuthal coordinate,
$\alpha_6 \in [0, \pi]$, and $\vec X$ cover a five sphere. We can look at
this sphere as immersed in $\mathbb C^3$ via
\eqn
&& \vec X =(z_1 \ , z_2 \ , z_3 ) =
\left( \cos \alpha_5 e^{i\alpha_4} \ ,
\sin \alpha_5 \cos \alpha_2
e^{i\left( \alpha_1 +\alpha_3 +\frac {\alpha_4}2\right)} \ ,
\sin \alpha_5 \sin \alpha_2
e^{i\left( \alpha_1 -\alpha_3 -\frac {\alpha_4}2\right)} \right) \ , \cr
&& \alpha_1 \in \left[ 0\ , \pi \right] \ , \qquad
\alpha_2 \in \left[ 0\ , \frac \pi2 \right] \ , \qquad
\alpha_3 \in \left[ 0\ , 2\pi \right] \ , \qquad
\alpha_4 \in \left[ 0\ , 2\pi \right] \ , \qquad
\alpha_5 \in \left[ 0\ , \frac \pi2 \right]  \ . \nonumber
\feqn
Computing the metric $ds^2_{S^5} =|dz_1|^2 +|dz_2|^2 +|dz_3|^2$ in these
coordinates we find
\eqn
\frac 43 ds^2_{BASE} =d\alpha_6^2 +\sin^2 \alpha_6
\left\{ ds^2_{S^5}  \right\} \ .
\feqn
This complete our identification.

%%%%%%%%%%%%%%%%%%%%%%%%%%%%%%%%%%%%%%%%%%%%%%%%%%%%%%%%%%%%%%%%%%%%%%%%%%%%%%%%%%%%%%%%%%%%%%%%%%%%%%%%%%%%%%%

\section{Conclusions}
\label{finrem}

We can thus parametrize the elements of the group $G_2$ as
\eqn
g=
e^{\alpha_1 C_3} e^{\alpha_2 C_2} e^{\alpha_3 C_3} e^{\frac {\sqrt 3}2 \alpha_4 C_8}
e^{\alpha_5 C_5} e^{\frac {\sqrt 3}2 \alpha_6 C_9}
SU(3)[\gamma_1 \ , \ldots \ ,\gamma_8]
\feqn
with
\eqn
&& \alpha_1 \in \left[ 0\ , \pi \right] \ , \qquad
\alpha_2 \in \left[ 0\ , \frac \pi2 \right] \ , \qquad
\alpha_3 \in \left[ 0\ , 2\pi \right] \ , \qquad
\alpha_4 \in \left[ 0\ , 2\pi \right] \ , \cr
&& \alpha_5 \in \left[ 0\ , \frac \pi2 \right] \ , \qquad
\alpha_6 \in \left[ 0\ , \pi \right] \ , \qquad
\gamma_1 \in \left[ 0\ , 2\pi \right] \ , \qquad
\gamma_2 \in \left[ 0\ , \frac \pi2 \right] \ , \cr
&& \gamma_3 \in \left[ 0\ , \pi \right] \ , \qquad
\gamma_4 \in \left[ 0\ , \frac \pi2 \right] \ , \qquad
\gamma_5 \in \left[ 0\ , 2\pi \right] \ , \qquad
\gamma_6 \in \left[ 0\ , \pi \right] \ , \cr
&& \gamma_7 \in \left[ 0\ , \frac \pi2 \right] \ , \qquad
\gamma_8 \in \left[ 0\ , \pi \right] \ .
\feqn
From (\ref{metric}) one could easily find the biinvariant metric.
The corresponding invariant measure is
\eqn
d\mu_{G_2} = \frac {27}{32} \sin^5 \alpha_6 \cos \alpha_5 \sin^3 \alpha_5
\sin (2\alpha_2) d\mu_{SU(3)} d\alpha_6 d \alpha_5 d\alpha_4 d \alpha_3
d\alpha_2 d \alpha_1 \ .
\feqn
$d\mu_{SU(3)}$ being the invariant measure over $SU(3)$ as given in App.\ref{app:su(3)}.\\
This is a simple parametrization for $G_2$, which
underlines the structure of $G_2$ as a $SU(3)$ fibration  over $S^6$. It could be used to
implement analytical or numerical computations in lattice gauge theory and in random
matrix models.\\
Here we used a geometrical approach to determine the range of parameters, but
the topological method of \cite{Cacciatori:2005yb} could also be used, giving the same results.
%%%%%%%%%%%%%%%%%%%%%%%%%%%%%%%%%%%%%%%%%%%%%%%%%%%%%%%%%%%%%%%%%%%%%%%%%%%%%%%%%%%%%%%%%%%%%%%%%%%%%%%%%%%%%%%

\section*{Acknowledgments}
\small

I would like to thank Professor Ruggero Ferrari which introduced me to the problem, 
Bianca Letizia Cerchiai for computer checks,
Professor Antonio Scotti for his participation and my friend Aldo Cleric\`o for
encouragements.\\
I also thank Dietmar Klemm, Alberto Della Vedova, Giovanni Ortenzi and Giuseppe Berrino 
for useful discussions.\\
This work was partially supported by INFN, COFIN prot. 2003023852\_008 and the 
European Commission RTN program MRTN--CT--2004--005104 in which S.~C.~is
associated to the University of Milano--Bicocca.
\normalsize

\newpage

\begin{appendix}

%%%%%%%%%%%%%%%%%%%%%%%%%%%%%%%%%%%%%%%%%%%%%%%%%%%%%%%%%%%%%%%%%%%%%%%%%%%%%%%%%%%%%%%%%%%%%%%%%%%%%
\section{Euler parametrization for \boldmath{$SU(3)$}.}
\label{app:su(3)}
The Euler parametrization for $SU(3)$ can be easily obtained as\footnote{again the ranges could be found using
the topological method, however we simply adapted the results shown in App. B of \cite{Cvetic:2001zx} to our case.} 
\eqn
SU(3)[\gamma_1, \ldots ,\gamma_8] =e^{\gamma_1 C_3} e^{\gamma_2 C_2}e^{\gamma_3 C_3}
e^{\gamma_4 C_5}e^{\sqrt 3 \gamma_5 C_8}e^{\gamma_6 C_3}e^{\gamma_7 C_2}e^{\gamma_8 C_3} \ , 
\feqn
with range
\eqn
&& \gamma_1 \in \left[ 0\ , 2\pi \right] \ , \qquad
\gamma_2 \in \left[ 0\ , \frac \pi2 \right] \ , \qquad
\gamma_3 \in \left[ 0\ , \pi \right] \ , \qquad
\gamma_4 \in \left[ 0\ , \frac \pi2 \right] \ , \cr
&& \gamma_5 \in \left[ 0\ , 2\pi \right] \ , \qquad
\gamma_6 \in \left[ 0\ , \pi \right] \ , \qquad
\gamma_7 \in \left[ 0\ , \frac \pi2 \right] \ , \qquad
\gamma_8 \in \left[ 0\ , \pi \right] \ .
\feqn
In this way ${\gamma_1 , \gamma_2 , \gamma_3 }$ cover an $SU(2)$ subgroup, $\gamma_5$
covers $U(1)$ and ${\gamma_6 , \gamma_7 , \gamma_8 }$ cover $SO(3)$.\\
The resulting invariant measure is
\eqn
d\mu_{SU(3)} =\sqrt 3 \sin (2\gamma_2) \sin^3 \gamma_4 \cos \gamma_4 \sin (2\gamma_7) \prod_{i=1}^8 d\gamma_i .
\feqn
%%%%%%%%%%%%%%%%%%%%%%%%%%%%%%%%%%%%%%%%%%%%%%%%%%%%%%%%%%%%%%%%%%%%%%%%%%%
\section{\boldmath{$J_h$} currents.}
\label{app:curr}
Here we give the currents $J_h$ used to generate the metric on the base manifold. The relations we need follow
from the commutators matrix given in App. \ref{app:comm} and are
\eqn
&& e^{-xC_3}C_2 e^{xC_3} =\cos (2x) C_2 +\sin (2x) C_1 \cr
&& e^{-xC_3}C_1 e^{xC_3} =\cos (2x) C_1 -\sin (2x) C_2 \cr
&& e^{-xC_2}C_3 e^{xC_2} =\cos (2x) C_3 -\sin (2x) C_1 \cr
&& e^{-xC_5}C_1 e^{xC_5} =\cos x C_1 +\sin x C_6 \cr
&& e^{-xC_5}C_2 e^{xC_5} =\cos x C_2 +\sin x C_7 \cr
&& e^{-xC_5}C_3 e^{xC_5} =\frac 14 (3+\cos (2x))C_3 -\frac 12 \sin (2x) C_4 -\frac {\sqrt 3}2 \sin^2 x C_8 \cr
&& e^{-xC_5}C_8 e^{xC_5} =\frac 14 (1+3\cos (2x))C_8 -\frac {\sqrt 3}2 \sin (2x) C_4 -\frac {\sqrt 3}2 \sin^2 x C_3 \cr
&& e^{-\sqrt 3 xC_9}C_4 e^{\sqrt 3 xC_9} =\cos^3 x C_4 -\sin^3 x C_7 +\sqrt 3 \cos x \sin^2 x C_{11} -\sqrt 3 \sin x \cos^2 x C_{14} \cr
&& e^{-\sqrt 3 xC_9}C_5 e^{\sqrt 3 xC_9} =\cos^3 x C_5 +\sin^3 x C_6 +\sqrt 3 \cos x \sin^2 x C_{12} +\sqrt 3 \sin x \cos^2 x C_{13} \cr
&& e^{-\sqrt 3 xC_9}C_6 e^{\sqrt 3 xC_9} =\cos^3 x C_6 -\sin^3 x C_5 +\sqrt 3 \cos x \sin^2 x C_{13} -\sqrt 3 \sin x \cos^2 x C_{12} \cr
&& e^{-\sqrt 3 xC_9}C_7 e^{\sqrt 3 xC_9} =\cos^3 x C_7 +\sin^3 x C_4 +\sqrt 3 \cos x \sin^2 x C_{14} +\sqrt 3 \sin x \cos^2 x C_{11} \cr
&& e^{-\sqrt 3 xC_9}C_8 e^{\sqrt 3 xC_9} =\cos (2x) C_8 +\sin (2x) C_{10} \ ,
\feqn
and the fact that $C_9$ commutes with $C_1 , C_2$ and $C_3$. \\
If we put
\eqn
&& s_1 =-\sin (2\alpha_2) \cos (2\alpha_3) d\alpha_1 +\sin(2\alpha_3)d\alpha_2 \cr
&& s_2 =\sin (2\alpha_2) \sin (2\alpha_3) d\alpha_1 +\cos(2\alpha_3)d\alpha_2 \cr
&& s_3 =\cos (2\alpha_2) d\alpha_1 +d\alpha_3
\feqn
the resulting currents are then
\eqn
&& J_h^1 =\cos \alpha_5 s_1 \cr
&& J_h^2 =\cos \alpha_5 s_2 \cr
&& J_h^3 =\frac 14 (3+\cos (2\alpha_5 )) s_3 -\frac 34 \sin^2 \alpha_5 d\alpha_4 \cr
&& J_h^4 =-\frac 12 \sin (2\alpha_5 )\cos^3 \frac {\alpha_6}2  \left[s_3 +\frac 32 d\alpha_4 \right] +\sin a_5 \sin^3 \frac {\alpha_6}2 s_2 \cr
&& J_h^5 =-\sin \alpha_5 \sin^3 \frac {\alpha_6}2 s_1 +\cos^3 \frac {\alpha_6}2 d\alpha_5 \cr
&& J_h^6 =\sin \alpha_5 \cos^3 \frac {\alpha_6}2 s_1 +\sin^3 \frac {\alpha_6}2 d\alpha_5 \cr
&& J_h^7 =\frac 12 \sin (2\alpha_5 )\sin^3 \frac {\alpha_6}2  \left[s_3 +\frac 32 d\alpha_4 \right] +\sin \alpha_5 \cos^3 \frac {\alpha_6}2 s_2 \cr
&& J_h^8 =\frac {\sqrt 3}2 \cos \alpha_6  \left[ \frac 14 (1+3\cos (2\alpha_5 ))d\alpha_4 - \sin^2 \alpha_5 s_3 \right] \cr
&& J_h^9 =\frac {\sqrt 3}2 d\alpha_6 \cr
&& J_h^{10} =\frac {\sqrt 3}2 \sin \alpha_6 \left[\frac 14 (1+3\cos (2\alpha_5 ))d\alpha_4 -\sin^2 \alpha_5 s_3 \right] \cr
&& J_h^{11} =-\frac {\sqrt 3}2 \sin (2\alpha_5 ) \cos \frac {\alpha_6}2 \sin^2 \frac {\alpha_6}2 \left[ s_3 +\frac 32 d\alpha_4 \right]
+\sqrt 3 \sin \alpha_5 \sin \frac {\alpha_6}2 \cos^2 \frac {\alpha_6}2 s_2 \cr
&& J_h^{12} =-\sqrt 3 \sin \alpha_5 \sin \frac {\alpha_6}2 \cos^2 \frac {\alpha_6}2 s_1 
+\sqrt 3 \cos \frac {\alpha_6}2 \sin^2 \frac {\alpha_6}2 d\alpha_5 \cr
&& J_h^{13} =\sqrt 3 \sin \alpha_5 \cos \frac {\alpha_6}2 \sin^2 \frac {\alpha_6}2 s_1 
+\sqrt 3 \sin \frac {\alpha_6}2 \cos^2 \frac {\alpha_6}2 d\alpha_5 \cr
&& J_h^{14} =\frac {\sqrt 3}2 \sin (2\alpha_5 ) \sin \frac {\alpha_6}2 \cos^2 \frac {\alpha_6}2 \left[ s_3 +\frac 32 d\alpha_4 \right]
+\sqrt 3 \sin \alpha_5 \cos \frac {\alpha_6}2 \sin^2 \frac {\alpha_6}2 s_2 \ .
\feqn
These currents together the $SU(3)$ currents (as given in \cite{Cvetic:2001zx}) can be used in (\ref{metric}) to compute
the full (bi-)invariant metric of $G_2$.
%%%%%%%%%%%%%%%%%%%%%%%%%%%%%%%%%%%%%%%%%%%%%%%%%%%%%%%%%%%%%%%%%%%%%%%%%%%%%%%%%%%%%%%%%%%%%%%%%%%%%
\begin{sidewaystable}
\centering
\section{The commutators matrix.}
\label{app:comm}
{\tiny
\vspace{3cm}
%\begin{picture}
%\begin{sidewaystable}
\begin{equation*}
\begin{array}{rcl}
B=\left(
\begin{array}{cccccccccccccc}
0 & 2C_3 & -2C_2 & C_7 & C_6 & -C_5 & -C_4 & 0 & 0 & 0 & C_{14} & C_{13} & -C_{12} & -C_{11} \\
* & 0 & 2C_1 & -C_6 & C_7 & C_4 & -C_5 & 0 & 0 & 0 & -C_{13} & C_{14} & C_{11} & -C_{12} \\
* & * & 0 & C_5 & -C_4 & C_7 & -C_6 & 0 & 0 & 0 & C_{12} & -C_{11} & C_{14} & -C_{13} \\
* & * & * & 0 & C_3 +\sqrt 3 C_8 & -C_2 & C_1 & -\sqrt 3 C_5 & -C_{14} & -C_{13} & 0 & 0 & C_{10} & C_9 \\
* & * & * & * & 0 & C_1 & C_2 & \sqrt 3 C_4 & C_{13} & -C_{14} & 0 & 0 & -C_9 & C_{10} \\
* & * & * & * & * & 0 & C_3 -\sqrt 3 C_8 & \sqrt 3 C_7 & -C_{12} & C_{11} & -C_{10} & C_9 & 0 & 0 \\
* & * & * & * & * & * & 0 & -\sqrt 3 C_6 & C_{11} & C_{12} & -C_9 & -C_{10} & 0 & 0 \\
* & * & * & * & * & * & * & 0 & \frac 2{\sqrt 3} C_{10} & -\frac 2{\sqrt 3}
C_9 & -\frac 1{\sqrt 3} C_{12} & \frac 1{\sqrt 3} C_{11} & \frac 1{\sqrt 3}
C_{14} & -\frac 1{\sqrt 3} C_{13} \\
* & * & * & * & * & * & * & * & 0 & \frac 2{\sqrt 3} C_8 & C_7 -\frac 2{\sqrt 3}
C_{14} & \frac 2{\sqrt 3} C_{13} -C_6 & C_5 -\frac 2{\sqrt 3} C_{12} &
\frac 2{\sqrt 3} C_{11} -C_4 \\
* & * & * & * & * & * & * & * & * & 0 & \frac 2{\sqrt 3} C_{13} +C_6 &
\frac 2{\sqrt 3} C_{14} +C_7 & -\frac 2{\sqrt 3} C_{11} -C_4 &
-\frac 2{\sqrt 3} C_{12} +C_5 \\
* & * & * & * & * & * & * & * & * & * & 0 &
-\frac 1{\sqrt 3} C_{8} +C_3 & \frac 2{\sqrt 3} C_{10} -C_2 &
-\frac 2{\sqrt 3} C_{9} +C_1 \\
* & * & * & * & * & * & * & * & * & * & * & 0 & \frac 2{\sqrt 3} C_{9} +C_1 & \frac 2{\sqrt 3} C_{10} +C_2 \\
* & * & * & * & * & * & * & * & * & * & * & * & 0 & \frac 1{\sqrt 3} C_{8} +C_3 \\
* & * & * & * & * & * & * & * & * & * & * & * & * & 0
\end{array}
\right)
\end{array}
\end{equation*}
}
\end{sidewaystable}
\end{appendix}

\end{document}